# Future-proofing Education: A Prototype for Simulating Oral Examinations Using Large Language Models

André Nitze
Department of Business
Brandenburg University of Applied Sciences
Brandenburg an der Havel, Germany
andre.nitze@th-brandenburg.de

*Abstract*—This study explores the impact of Large Language Models (LLMs) in higher education, focusing on an automated oral examination simulation using a prototype. The design considerations of the prototype are described, and the system is evaluated with a select group of educators and students. Technical and pedagogical observations are discussed. The prototype proved to be effective in simulating oral exams, providing personalized feedback, and streamlining educators' workloads. The promising results of the prototype show the potential for LLMs in democratizing education, inclusion of diverse student populations, and improvement of teaching quality and efficiency.

*Keywords—Artificial Intelligence, Large Language Models, Inclusive Education, Oral Examinations, Simulations, STEM Pedagogy.*

## I. INTRODUCTION

The 21st-century STEM education landscape is rapidly transforming, with advanced information technologies, particularly Artificial Intelligence (AI) and Large Language Models (LLMs), playing a pivotal role in this evolution. An introduction and overview on LLMs can be found in [1]. As the educational sector grapples with the implications of AI-generated texts, oral examination conversations stand the test of time as a direct and straightforward method for assessing students, ensuring genuine comprehension and knowledge application beyond generated content.

The advent of AI and LLMs in education has sparked a paradigm shift which challenges traditional teaching methodologies. Owing to their increased dissemination in academia, LLMs are being used to generate multiple-choice questions, provide individualized feedback, and perform or support various other teaching tasks. The models are trained on vast amounts of text data, allowing them to generate questions that align with the desired learning outcomes. By leveraging LLMs, educators can save time and effort in creating assessments, and focus more on instructional design and personalized teaching. Furthermore, LLMs can assist in generating explanations, providing real-time feedback, and providing additional learning resources. The versatility of LLMs makes them a valuable tool for enhancing traditional teaching practices and adapting to the evolving educational needs.

With the announcement of customizable generative pre-trained transformers (GPTs) by OpenAI [2], several new features were introduced, which were previously difficult to implement by software engineers, let alone educators. These features included the use of custom text content and instructing a so-called "assistant" to answer questions based on a predefined persona. One of the advertised examples was a "study buddy", which inspired this work. Since the effectiveness of knowledge recall for long-term retention ("test effect") is empirically well documented [3], it is an obvious approach to apply the technology to create such recall situations in an automated fashion and help students prepare for oral examinations.

### A. Oral exams: Assessment solution for the "AI era"?

Oral examinations in higher education offer a unique and comprehensive method of assessing student learning, providing several advantages over traditional written assessments. This is due to their ability to evaluate a range of skills and competencies that are essential in today's rapidly evolving academic and professional landscapes.

Firstly, oral exams test the depth of understanding and critical thinking skills in a way that written tests may not. Students must articulate their knowledge of complex programming concepts, algorithms, and system designs in real-time. This format demands a solid grasp of the subject matter, as students cannot rely on memorized code or concepts. For instance, they may be asked to explain the workings of a specific algorithm or the design of a software system, showcasing their ability to think and reason like real-world software developers or system administrators.

Also, oral exams are particularly relevant because they assess communication skills. As AI and technology increasingly automate technical tasks, the ability to effectively communicate complex ideas becomes a crucial skill. Oral exams require students to present information clearly and coherently, a skill that is highly valuable in professional environments where one must often explain complex concepts to non-experts.

Moreover, oral exams can assess interdisciplinary knowledge, which is vital in computer science where integration with fields like mathematics, electrical engineering, and even psychology (in the case of human-computer interaction) is frequent. The method encourages a holistic understanding rather than the compartmentalized knowledge often tested in written exams.

Additionally, oral exams minimize the risk of academic dishonesty, a growing concern in the era of readily accessible information and sophisticated AI tools. The face-to-face format of oral exams makes it difficult for students to use unauthorized aids or plagiarize responses, ensuring a more accurate assessment of their individual capabilities and knowledge.

Oral examinations can also be more inclusive for certain student populations. For instance, students with specific learning disabilities, such as dyslexia, who might struggle with written exams, can often better demonstrate their understanding and knowledge orally.

While traditional written assessments have their place, oral examinations offer a valuable method of assessing a wide range of competencies. As such, they are particularly well-suited to the demands and challenges of the AI era.

*B. Research objectives*

Based on the benefits of oral examinations, this study aims to gauge the capabilities of LLMs to simulate oral examinations. Specifically, it focuses on the design, development, and evaluation of an LLM-based prototype to simulate oral examinations in higher education. The objectives include:

1. Provide a personalized learning experience, bridging individual knowledge gaps and helping individual students prepare for oral exams.

2. Assess the prototype's potential to reduce educator workload and enhance teaching quality.

3. Evaluate the suitability of one of the most popular available LLM for assessing knowledge in higher education syllabi.

## II. RELATED WORK

The research regarding application of LLMs in preparing students for oral examinations is limited. However, LLMs are already being used to create educational content, improve student engagement and interaction, and personalize learning experiences as shown by Caines et al. [4]. These applications have demonstrated the potential of LLMs in transforming various aspects of educational practices.

Kasneci et al. describe how LLMs can improve curriculum development, teaching methodologies, and student assessments [5]. The researchers also point out the need for human oversight to cope with the models" biases and factual errors, emphasizing the importance of a balanced approach in integrating LLMs into educational settings.

Additionally, Abd-Alrazaq et al. demonstrated that LLMs can be utilized in automated feedback systems to provide detailed and natural-sounding feedback to students, helping them learn better [6]. This innovation has been instrumental in enhancing the quality and efficiency of educational feedback mechanisms. However, the integration of LLMs in education comes with challenges such as algorithmic bias, overreliance, plagiarism, misinformation, and privacy concerns (ibid). These issues highlight the need for cautious and responsible deployment of LLMs in educational contexts.

Dai et al. have used the widely known LLM "ChatGPT" to generate textual feedback on student assignments [7]. They found that not only is the feedback provided by the LLM more detailed than what a human instructor would have provided, but the human teachers also agreed with the model's assessment, and students" learning processes benefitted directly from the feedback.

These findings are promising regarding automated assessment and feedback using LLMs and suggest a significant potential for LLMs in augmenting traditional educational methods.

In conclusion, while the application of LLMs in preparing students for oral examinations remains underexplored, existing research provides a strong foundation for future studies. It is evident that LLMs hold considerable promise in enhancing educational practices, though their integration must be approached with careful consideration of the accompanying challenges and ethical implications.

## III. METHOD

A prototype was developed to assess the feasibility of using LLMs to simulate oral exams in STEM education. The following sections describe the development process of this system.

*A. Requirements gathering*

The requirements for the prototype were primarily derived from valuable insights and experiences gained as educators. Drawing on this knowledge, the aim was to develop a solution that addresses the specific needs and challenges faced by students, with a focus on enhancing their learning experiences. The requirements include a personalized level of difficulty, personalized feedback, workload reduction for educators, and provision of an inclusive educational environment for students with diverse backgrounds, knowledge levels, and native languages. By leveraging firsthand expertise in the field, the prototype aimed to provide tangible benefits to students and educators.

*B. Software architecture and user interface design*

Fig. 1 shows the software architecture. The software is split into front- and back-end components. The front-end was built using Typescript and the "VueJS" framework, providing a user-friendly interface for educators and students. The back-end was developed using Python and the "Flask" framework, handling the server-side communication with the third-party REST API (Application Programming Interface) and providing a similar REST-based API to the front-end. The back-end accesses the OpenAI REST API. Calls to the API are made from the back-end using the endpoints and authentication methods specified in the OpenAI documentation.

During the design phase, it was also important to create a familiar interface that resembles a normal chat to reduce entry barriers for users. In addition to a display of the chat history, the user interface consists of a text input field, a "send" button and an optional "Give hint" button.

*C. Creating the assistant*

The prototype uses OpenAI's "Assistants API" to submit user messages and generate matching responses. The assistant leverages existing language models, like "GPT-3.5-Turbo" or "GPT-4". Comprehensive instructions were devised as initial prompt to the assistant. These instructions

define greatly how the LLM will "behave" towards the users. The instructions included the following aspects:

- Role: Tutor preparing students for oral exams.
- Process: Ask a question within a specified subject area and let students respond.
- Feedback: Provide detailed, subject-specific feedback on each response (e.g., quality, completeness, correctness, precision).
- Identify and correct misinformation and ask follow-up questions for unclear responses.
- Utilize slide sets provided by the user or system, if available.
- Start with simple questions to assess students" knowledge level, then progressively ask harder questions.
- Provide a rating or grade upon request or after five answered questions, using the university's grading scale (1.0, 1.3, 1.7, 2.0 up to 4.0, and 5.0) and as a percentage (0-100%).
- Ask if the student wishes to continue, then proceed with the same or a new topic, or conclude the session.
- Point out that the rating applies only to the discussed subject area.
- For a manually requested evaluation, at least three questions must have been answered.
- Offer hints if students struggle to answer a question.
- Present questions in a concrete application example if possible.
- Mention that proficiency in the exam language may affect the grade if students use other languages to prepare for the exam.
- Respond with a small hint upon receiving the exact message "%REQUEST_HINT%".

While technically feasible, an upload option for users to provide custom files was not implemented in the prototype. However, two PDF files were added to the assistant manually. One file contained unedited lecture slides from one specific course unit to test the assistant's ability to only assess the topics covered within the slide deck. The other file contained made-up laws of a fictitious country to test the assistant's ability to retrieve knowledge for which it could not itself have a source in the training data.

*D. Testing*

To gather insights and feedback, a limited testing phase was conducted. A select group of educators and students present at a teaching-related event on the author's university campus was invited to evaluate the prototype. The goal was to assess the prototype's general usability, effectiveness, and potential impact on the teaching and learning experience.

*E. Deployment*

Once the prototype reached a stable state, it was deployed to a secure hosting environment, making it accessible to a limited group of educators and students for testing and evaluation.

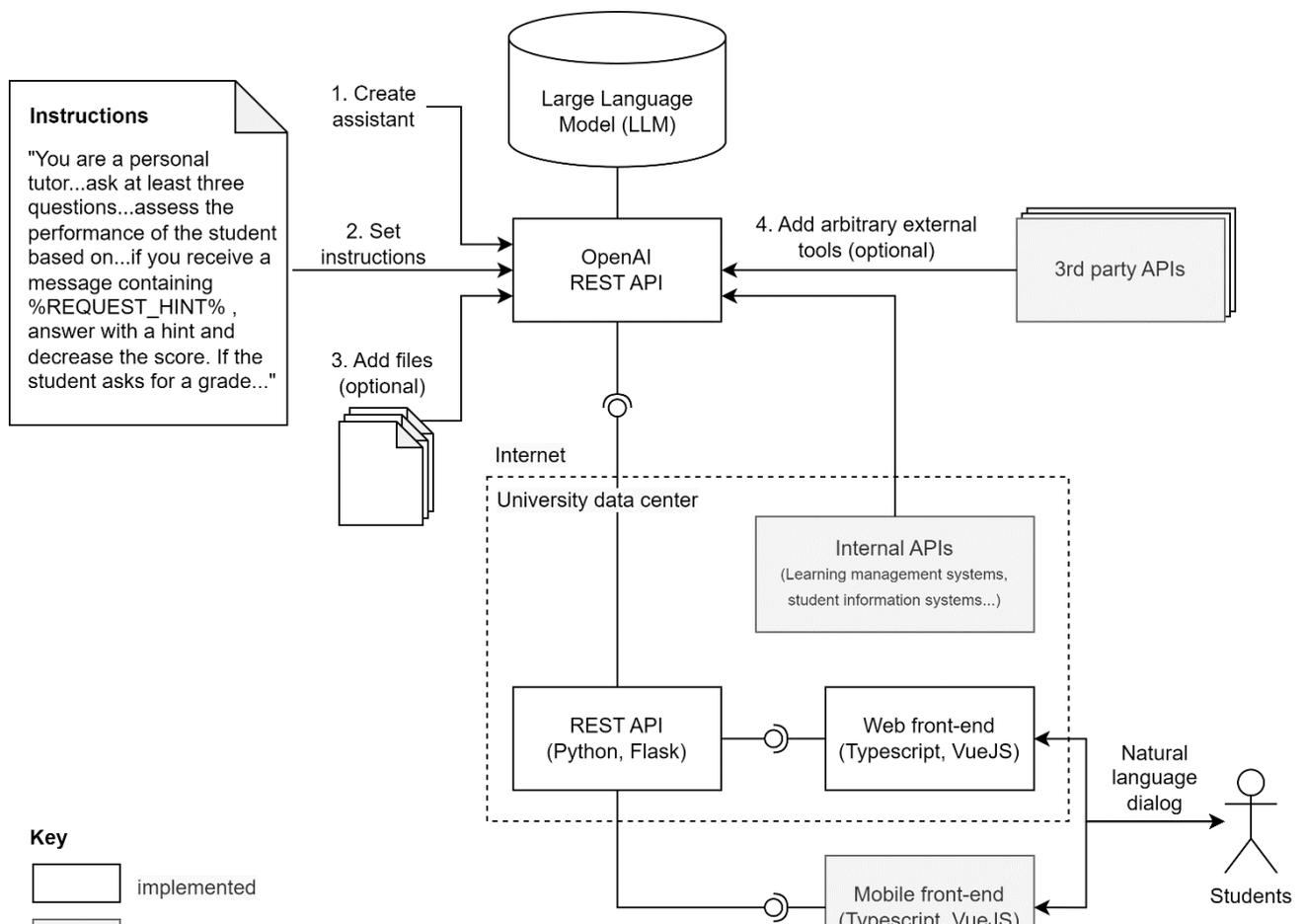

Fig. 1. Software architecture of the prototype with front-end and back-end components, and third-party API.

## IV. RESULTS

The following sections describe the results of the study from different perspectives, particularly from the pedagogical (A), technical (B), personal (C), and privacy and data protection (D) perspectives.

### A. Pedagocical observations

From a pedagogical perspective, the assistant acts as a reliable partner for asking questions about all tested topics and providing correct answers that help students prepare for oral exams. The assistant behaves as previously defined and thus offers a sparring partner for learners of all disciplines. It adapts to the language entered by the learner and answers and poses questions on individual requested topics. It follows up on unclear points and gives tips for the oral exam situation. Finally, the system awards a grade on the defined scale as required.

While the evaluation was conducted mostly within the field of computer science and business information systems, there's no reason to assume that other disciplines would be in any way less suited. The range and depth of covered topics of the utilized LLM is immense.

The occurrence of factual errors has been low overall. This is attributed to the relatively advanced model (GPT-4-preview) and the usual limitations of an oral exam in terms of length of questions.

Also, the requirement to reason about the logic of requests is low in this specific application. Furthermore, instructions can be tailored ad-hoc, i.e., while the assistant is being used. While testing, some skeptic educators have tried to trip up the assistant with trick questions, e.g., by stating a question containing a wrong assumption. The LLM usually handled such requests properly by pointing out the "flawed" question or just ignored this nuance of the request.

The actual feedback on given answers was precise. Textbook-quality definitions were recognized as higher quality by the LLM (*"Your answer was structured well, with clear sections for the definitions, components, isolation versus efficiency, and management practices. If this were included in an oral exam context, you would demonstrate a strong grasp of key operating systems concepts."*).

Likewise, low-quality, or vague answers are pointed out (*"Your response indicates you understand the broad role of an operating system, but in a formal exam setting, expanding on these points with specific examples or details would likely be expected."*). This allows the students to deepen their knowledge by either directly talking to the assistant or using other learning materials.

Finally, the accuracy and reliability of the simulated oral exams depends heavily on the capabilities of the LLMs used. This means that factual errors and false assertions can occur in the conversations. These errors cannot be corrected as there is no human supervision. Although errors can also occur in regular exam preparation with lecturers and other students, this possibility must be explicitly communicated to potential future users.

### B. Technical observations

From a technical perspective it is noteworthy that the model's response behavior can be precisely tailored to a specific niche while still utilizing the broader capabilities of the entire model by just providing the appropriate instruction in written form. That means, it contains the world's knowledge about any topic that is covered in most universities" syllabi and is still able to focus on helping students prepare for a narrow subject's exam. While these capabilities of LLMs have been known for some time and can be considered an aspect of the more familiar "prompt engineering", the aspect of "programming behavior through natural language text" opens a whole new set of possibilities.

Furthermore, the ability to feed individual content such as portable document format (PDF) files, formatted text documents (e.g., DOCX, PPTX), and plain text files (TXT, CSV) into the assistant improves its capabilities while keeping implementation effort low. One of the key limitations of previous versions of OpenAI's ChatGPT has been the limited context window. The context window describes how many tokens (parts of words) can be used in a prompt. Although the context window has been increased over the course of several versions, the API never came close to providing support for full lecture scripts or textbooks without the use of vector databases and additional tooling. As this restriction no longer applies, individual content can be made available to the model in advance or even uploaded directly by users.

Generating answers takes some time (up to five seconds), which can feel like a lot for current generations of users. However, since the students also must formulate and enter answers themselves, the overall speed of the process is somewhat slower anyway.

A last technical aspect is the ability to provide implementation-specific hints to the assistant. This means that the instructions given to the assistant can be used in parsers in the front-end and back-end to modify the response or just change the style of the output. In the prototype, this is demonstrated by using the %REQUEST_HELP% string as a normal request in the dialog. The LLM recognizes the distinct label and reacts accordingly with a content-related hint for the student as can be seen in Fig. 3. The front-end can interpret that string and show an appropriate icon for the user. This capability leads to diverse possibilities for programmatic use within the front-end application, e.g., rendering the final score in a specific way or providing other pre-defined interactions to the user.

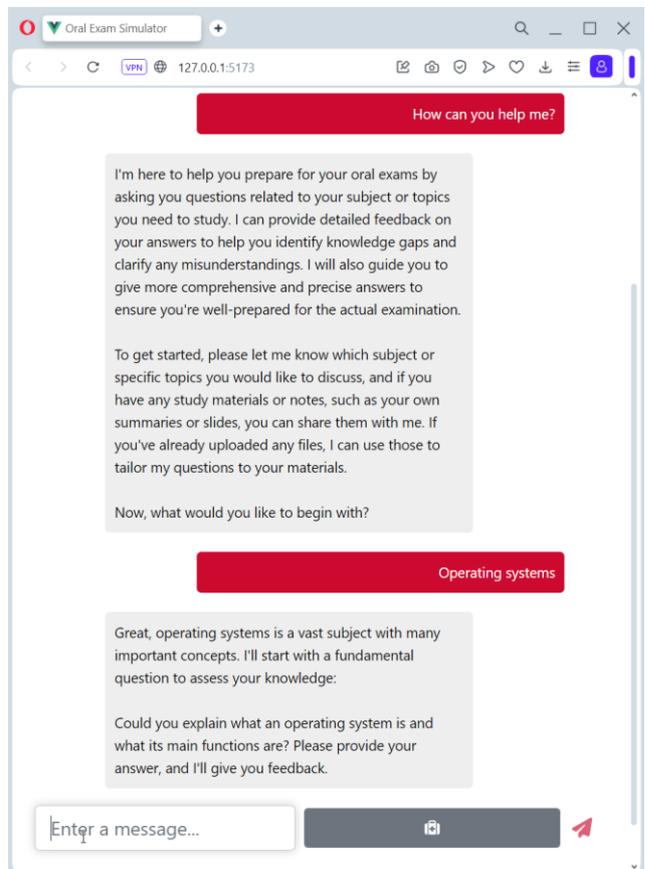

Fig. 2. User interface of the prototype with user messages (right), LLM-generated explanation of the assistant's purpose and an example question (left), and text input, "give hint" button, and submit button (bottom).

Despite the assistant's instructions, it is still easy to "convince" the model to do other things than intended ("LLM jailbreak"). This could lead to problems like leakage of training data and (costly) misuse of the assistant for other, possibly illegal, purposes. Further work is needed to make sure the scope of the assistant is limited to its" intended purpose or that requests per user are limited.

### C. Personal feedback from educators and students

During the test phase, a small group of interested teachers and students were offered the opportunity to try out the prototype for themselves. These trials were observed in person to obtain direct user feedback. The participants were provided with access to the prototype without much instruction above its" main task ("oral exam simulator").

Participants engaged with the prototype enthusiastically, providing valuable spontaneous feedback on their interactions and the responses generated by the LLM. Many even commented unprompted on their own answers and the responses of the LLM, as they also found themselves in a less serious, but nonetheless realistic examination situation with observers standing nearby.

Feedback from educators and students indicated several positive aspects of the prototype. Educators immediately recognized and appreciated the workload reduction provided by the automated feedback generation. All participants found the personalized feedback valuable and appreciated the interactive nature of the simulation. Participants who were not accustomed to using chatbots, as anticipated, have been overwhelmed by the precision of answers.

However, the testing phase also revealed some areas for improvement. Educators highlighted the need for more flexibility in customizing the examination scenarios and the importance of ensuring the accuracy and reliability of the LLM-generated responses. Students expressed the desire for additional features, such as real-time interaction with the prototype, especially by using a voice-based interface.

Overall, the limited testing phase already provided valuable insights and feedback, highlighting the potential of the prototype in enhancing the teaching and learning experience. It not only showed potential in offering personalized feedback but also in streamlining the preparation process for students and educators. Significantly, the prototype boasts high generalizability, necessitating minimal adaptations to accommodate diverse curricula.

### D. Privacy and data protection

While the use of a large-scale language model offers significant benefits in terms of accessibility to knowledge, careful consideration and implementation of privacy and data protection measures are crucial. These measures not only protect intellectual property and individual user data but also enhance the trustworthiness and reliability of the system, making it a more viable tool for educational and research purposes.

Since the assistant leverages the language model's huge training data set containing most of the world's knowledge, individual content like lecture slides or additional texts are optional. Hence, privacy of intellectual property is rarely an issue. If, for some reason, individual content needed to be included that required special data protection, an option would be to use an open-source LLM running locally in the university's data center. However, this would require a certain amount of computing capacity at the university's data center.

Furthermore, in a production-ready system, robust authentication and authorization mechanisms are imperative to safeguard user identities and access rights. These security measures ensure that only authorized users can access the system and their interactions remain confidential and system load can be controlled.

## V. CONCLUSION

This exploratory study, centered on the exploration and development of an LLM-based prototype for simulating oral examinations in higher education, has made notable strides in understanding, and harnessing the potential of LLMs in an educational context, and contributing to the broader discourse on the integration of AI in education. While the findings are preliminary, given the prototype's stage of development and the limited scope of validation, they offer valuable insights into the capabilities and limitations of LLMs in this novel application. The most recent version of the OpenAI API with its assistants has shown a remarkable capability to tailor learning experiences to diverse student populations.

Firstly, the developed prototype has shown promise in conducting straightforward knowledge assessments and individual feedback during the initial stages of academic pursuit. This aspect is particularly significant as it aligns with the evolving needs of personalized education, catering to the unique learning trajectories of students. The observations and direct user feedback indicate that the prototype is easily accepted as a learning tool. However, it's important to acknowledge that the depth and accuracy of personalization are depending on the prototype's current capabilities and require further refinement, e.g., more context information about the users, for comprehensive implementation.

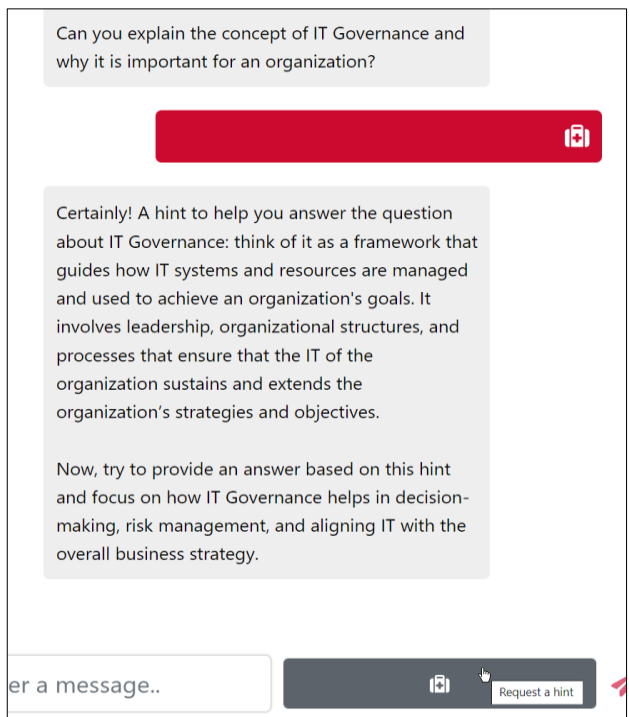

Fig. 3. "Request a hint" feature with custom rendering of the predefined %REQUEST_HINT% tag from the instructions.

Secondly, the study has begun to assess the potential of the prototype in reducing educator workload and enhancing teaching quality. Preliminary observations suggest that it can indeed assist educators by automating preparing students for oral exams. This aspect, however, demands a more extensive evaluation to fully understand its impact on the educational ecosystem, particularly in terms of quality assurance and pedagogical effectiveness.

Thirdly, the evaluation of one popular LLM for assessing knowledge in higher education syllabi has yielded encouraging results. The LLM demonstrated a competent level of understanding and interaction in line with higher education standards, at least at the undergraduate level. More complex question types like small case studies and other competency-based exam formats need further investigation.

The path ahead involves rigorous testing, refinement, and an ongoing dialogue between technological advancement and educational needs, ensuring that such innovations align ethically and effectively with the goals of higher education. More research and development are needed to enhance the sophistication of the prototype (s. "Future Work") and expand its validation.

The devised prototype offers promising avenues for the democratization of STEM education through LLMs. Its capability to provide tailored learning experiences, combined with its adaptability across various educational contexts, accentuates its potential relevance in contemporary academic settings.

## VI. Future Work

The feedback collected will be used to iterate and refine the prototype further, addressing the identified limitations and incorporating additional features based on the needs and preferences of educators and students.

While there already is a feature for interpreting code, another addition to the feature set of the OpenAI API is making dynamic external data available within responses through arbitrary external APIs. Assistants can call these so-called "tools" to enrich answers. In this scenario, this might involve incorporating calculation tasks, an area where LLMs are known to have limitations. Another option is to integrate the prototype directly into the learning management system (LMS) of the university as shown in the architecture diagram (s. Fig. 1). This can be done using simple "IFrames" or, depending on the specific LMS, a direct API integration into learning elements and contents of courses.

The prototype is inherently limited in functionality and production-readiness. Hence, the following features should be considered for further development:

1. the ability to upload custom content such as slide decks in PDF format,
2. integration of audio input and output to facilitate voice-based conversations with the assistant,
3. inclusion of parameters such as field of study, current term of the student, covered topics in the curriculum (as opposed to what the model assumes to be the content based on the provided topic), the student's preferences (e.g., harder questions), as well as other contextual factors that could improve the precision of answers,
4. enhancement of the answering speed by replacing the REST API by a streaming API (which has not yet been available as of writing this paper),
5. introduction of an "exam mode", where answers are not commented on or improved by the AI and the student's answers are used for assessment,
6. more complex application logic to accommodate for the above features and make it easy to navigate and use the system, and
7. a robust, scalable, and secure environment in which students can access the system, including authentication and authorization of users to prevent misuse while preserving privacy.

The obvious next step is to improve the prototype into a production-ready application, e.g., as part of a mobile university app or directly integrated into existing learning management systems. On this basis, data and further insights can be collected from real use in practice.